\documentclass[12pt]{article}

\usepackage{amsfonts, amsmath}

\newcommand{\be}{\begin{equation}} \newcommand{\ee}{\end{equation}}

\begin{document}
\title{Fundamental Length,Deformed Density Matrix and New View
 on the Black Hole Information Paradox} \thispagestyle{empty}

\author{A.E.Shalyt-Margolin\hspace{1.5mm}\thanks
{Fax (+375) 172 326075; e-mail: a.shalyt@mail.ru;alexm@hep.by}}
\date{}
\maketitle
 \vspace{-25pt}
{\footnotesize\noindent  National Center of Particles and High
Energy Physics, Bogdanovich Str. 153, Minsk 220040, Belarus\\
{\ttfamily{\footnotesize
}}

\rm\normalsize \vspace{0.5cm}

\begin{abstract}
{In this paper Quantum Mechanics with Fundamental Length is chosen
as Quantum Mechanics at Planck's scale. This is possible due to
the presence in the theory of General Uncertainty Relations (GUR).
Here Quantum Mechanics with Fundamental Length is obtained as a
deformation of Quantum Mechanics. The distinguishing feature of
the proposed approach in comparison with previous ones, lies on
the fact that here density matrix subjects to deformation whereas
so far commutators have been deformed. The density matrix obtained
by deformation of quantum-mechanical density one is named
throughout this paper density pro-matrix, which at low energy
limit turns to the density matrix. This transition corresponds to
non-unitary one from Quantum Mechanics with GUR to Quantum
mechanics. Below the implications of obtained results are
enumerated.New view on the Black Holes Information Paradox are
discussed}
\end{abstract}
\newpage
\section{Introduction}
In this paper Quantum Mechanics with Fundamental Length is chosen
as Quantum Mechanics at Planck's scale. This is possible due to
the presence in the theory of General Uncertainty Relations. Here
Quantum Mechanics with Fundamental Length is obtained as a
deformation of Quantum Mechanics. The distinguishing feature of
the proposed approach in comparison with previous ones, lies on
the fact that here density matrix subjects to deformation whereas
so far commutators have been deformed. The density matrix obtained
by deformation of quantum-mechanical density one is named
throughout this paper density pro-matrix. Within our approach two
main features of Quantum Mechanics are conserved: the
probabilistic interpretation of the theory and the well-known
measuring procedure corresponding to that interpretation.It was
shown also, inflationary model contains two different (unitary
non-equivalent) Quantum Mechanics: the first one describes nature
at the Planck scale and it is based on the GUR. The second one is
obtained as a limit transition from Planck scale to low energy one
and it is based on the Heisenberg uncertainty relations. The
interpretation of obtained results as well as their implications
are discussed below, in particular the new view on  the Black
Holes Information Paradox.

\section{General Uncertainty Relations and Fundamental Length}
Let's start considering the Heisenberg Uncertainty relation
(position-momentum) \cite{r1} :
\begin{equation}\label{U1}
\Delta x\geq\frac{\hbar}{\Delta p}.
\end{equation}
In the last 14-15 years a lot of papers were issued in which
authors using string theory \cite{r2}, gravitation \cite{r3},
Quantum theory of black holes \cite{r4} and other methods
\cite{r5} shown that Heisenberg Uncertainty relations should be
modified. In particular, a high energy addition have to appear

\begin{equation}\label{U2}
\triangle x\geq\frac{\hbar}{\triangle p}+\alpha
L_{p}^2\frac{\triangle p}{\hbar}.
\end{equation}
\\Where $L_{p}$ - the Planck length,
$L_{p}=\surd\frac{G\hbar}{c^3}\simeq1,6\;10^{-35}m$ and
 $\alpha > 0$ is a constant. In paper \cite{r3} was shown this
 constant can be chosen equal to 1. However, here we will use
 $\alpha$ as an arbitrary constant without any concrete value.
 The  inequality (\ref{U2}) is quadratic with respect to
$\triangle p$
\begin{equation}\label{U3}
\alpha L_{p}^2({\triangle p})^2-\hbar \triangle x \triangle p+
\hbar^2 \leq0
\end{equation}
and from it follows the fundamental length is
\begin{equation}\label{U4}
\triangle x_{min}=2\surd\alpha L_{p}
\end{equation}
Since further we are going to base only on the existence of
fundamental length it is necessary to point out this fact was
established not only from GUR. For instance, in
\cite{r6},\cite{r7} using an ideal experiment dealing with
gravitation field it was obtained the lower bound on limit length,
which was improved in \cite{r8} without GUR to an estimate of the
type $\sim L_{p}$. In what follows we will use the abbreviations
GUR for General Uncertainty Relations(\ref{U2}) and UR for
Heisenberg Uncertainty Relations (\ref{U2}) correspondingly.

\section{Density matrix and its deformation }
Let's consider in some detail the equation (\ref{U4}). Squaring it
left and right side, we obtain
\begin{equation}\label{U5}
(\overline{\Delta\widehat{X}^{2}})\geq 4\alpha L_{p}^{2 }
\end{equation}
or in terms of density matrix
\begin{equation}\label{U6}
Sp[(\rho \widehat{X}^2)-Sp^2(\rho \widehat{X}) ]\geq 4\alpha
L_{p}^{2 }>0
\end{equation}
where $\widehat{X}$ is the coordinate operator. Expression
(\ref{U6}) gives the measuring rule used in QM. However, in the
case considered here, in comparison with QM, the right part of
(\ref{U6}) cannot be done arbitrarily near to zero since it is
limited by $l^{2}_{min}>0$, where due to GUR $l_{min} \sim L_{p}$.

 Apparently, this may be due to
the fact that QMFL with GUR (\ref{U2}) is unitary non-equivalent
to  QM with UR. Actually, in QM the left-hand side of (\ref{U6})
can be chosen arbitrary closed to zero, whereas in QMFL this is
impossible. But if two theories are unitary equivalent then, the
form of their spurs should be retained. Besides, a more important
aspect is contributing to unitary non-equivalence of these two
theories: QMFL contains three fundamental constants (independent
parameters) $G$, $c$ and $\hbar$, whereas QM contains only one:
$\hbar$. Within an inflationary model (see \cite{r10}), QM is the
low-energy limit of QMFL (QMFL turns to QM) for the expansion of
the Universe. In this case, the second term in the right-hand side
of (\ref{U2}) vanishes and GUR turn to UR. A natural way for
studying QMFL is to consider this theory as a deformation of QM,
turning to QM at the low energy limit (during the expansion of the
Universe after the Big Bang). We will consider precisely this
option. However differing from authors of papers \cite{r4},
\cite{r5} and others, we do not deform commutators, but density
matrix, leaving at the same time the fundamental
quantum-mechanical measuring rule (\ref{U6}) without changes. Here
the following question may be formulated: how should be deformed
density matrix conserving quantum-mechanical measuring rules in
order to obtain self-consistent measuring procedure in QMFL? For
answering to the question we will use the R-procedure. For
starting let us to consider R-procedure both at the Planck's
energy scale and at the low-energy one. At the Planck's scale $a
\approx il_{min}$ or $a \sim iL_{p}$, where $i$ is a small
quantity. Further $a$ tends to infinity and we obtain for density
matrix $$Sp[\rho a^{2}]-Sp[\rho a]Sp[\rho a] \simeq
l^{2}_{min}\;\; or\;\; Sp[\rho]-Sp^{2}[\rho] \simeq
l^{2}_{min}/a^{2}.$$

 Therefore:

 \begin{enumerate}
 \item When $a < \infty$, $Sp[\rho] =
Sp[\rho(a)]$ and
 $Sp[\rho]-Sp^{2}[\rho]>0$. Then, \newline $Sp[\rho]<1$
 that corresponds to the QMFL case.
\item When $a = \infty$, $Sp[\rho]$ does not depend on $a$ and
$Sp[\rho]-Sp^{2}[\rho]\rightarrow 0$. Then, $Sp[\rho]=1$ that
corresponds to the QM case.
\end{enumerate}
How should be points 1 and 2 interpreted? How does analysis
above-given agree to the main result from \cite{r18} \footnote
{"... there cannot be any physical state which is a position
eigenstate since a eigenstate would of course have zero
uncertainty in position"}? It is in full agreement. Indeed, when
state-vector reduction (R-procedure) takes place in QM then,
always an eigenstate (value) is chosen exactly. In other words,
the probability is equal to 1. As it was pointed out in the
above-mentioned point 1 the situation changes when we consider
QMFL: it is impossible to measure coordinates exactly since it
never will be absolutely reliable. We obtain in all cases a
probability less than 1 ($Sp[\rho]=p<1$). In other words, any
R-procedure in QMFL leads to an eigenvalue, but only with a
probability less than 1. This probability is as near to 1 as far
the difference between measuring scale $a$ and $l_{min}$ is
growing, or in other words, when the second term in (\ref{U2})
becomes insignificant and we turn to QM. Here there is not a
contradiction with \cite{r18}. In QMFL there are not exact
coordinate eigenstates (values) as well as there are not pure
states. In this paper we do not consider operator properties in
QMFL as it was done in \cite{r18} but density-matrix properties.

 The  properties of density matrix in
QMFL and QM have to be different. The only reasoning in this case
may be as follows: QMFL must differ from QM, but in such a way
that in the low-energy limit a density matrix in QMFL must
coincide with the density matrix in QM. That is to say, QMFL is a
deformation of QM and the parameter of deformation depends on the
measuring scale. This means that in QMFL $\rho=\rho(x)$, where $x$
is the scale, and for $x\rightarrow\infty$  $\rho(x) \rightarrow
\widehat{\rho}$, where $\widehat{\rho}$ is the density matrix in
QM.

Since on the Planck's scale $Sp[\rho]<1$, then for such scales
$\rho=\rho(x)$, where $x$ is the scale, is not a density matrix as
it is generally defined in QM. On Planck's scale we name $\rho(x)$
 "density pro-matrix". As follows from the above, the density
matrix $\widehat{\rho}$ appears in the limit
\begin{equation}\label{U12}
\lim\limits_{x\rightarrow\infty}\rho(x)\rightarrow\widehat{\rho},
\end{equation}
when GUR (\ref{U2}) turn to UR  and QMFL turns to QM.

Thus, on Planck's scale the density matrix is inadequate to obtain
all information about the mean values of operators. A "deformed"
density matrix (or pro-matrix) $\rho(x)$ with $Sp[\rho]<1$ has to
be introduced because a missing part of information $1-Sp[\rho]$
is encoded in the quantity $l^{2}_{min}/a^{2}$, whose specific
weight decreases as the scale $a$ expressed in  units of $l_{min}$
is going up.

\section{QMFL as a deformation of QM by density matrix}
Here we are going to describe QMFL as a deformation of QM using
the density pro-matrix formalism. In this context density
pro-matrix has to be understood as a deformed density matrix in
QMFL. As fundamental deformation parameter we will use
$\beta=l_{min}^{2}/x^{2 }$, where $x$ is the scale.

\noindent {\bf Definition 1.}

\noindent Any system in QMFL is described by a density pro-matrix
$\rho(\beta)=\sum_{i}\omega_{i}(\beta)|i><i|$, where
\begin{enumerate}
\item $0<\beta\leq1/4$;
\item The vectors $|i>$ form a full orthonormal system;
\item $\omega_{i}(\beta)\geq 0$ and for all $i$ there is a
finite limit $\lim\limits_{\beta\rightarrow
0}\omega_{i}(\beta)=\omega_{i}$;
\item
$Sp[\rho(\beta)]=\sum_{i}\omega_{i}(\beta)<1$,
$\sum_{i}\omega_{i}=1$;
\item For any operator $B$ and any $\beta$ there is a
 mean operator $B$, which depends on  $\beta$:\\
$$<B>_{\beta}=\sum_{i}\omega_{i}(\beta)<i|B|i>.$$
\end{enumerate}
At last, in order to match our definition with the result of
section 2 the next condition has to be fulfilled:
\begin{equation}\label{U13}
Sp[\rho(\beta)]-Sp^{2}[\rho(\beta)]\approx\beta,
\end{equation}
from which we can find the meaning of the quantity
$Sp[\rho(\beta)]$, which satisfies the condition of definition:
\begin{equation}\label{U14}
Sp[\rho(\beta)]\approx\frac{1}{2}+\sqrt{\frac{1}{4}-\beta}.
\end{equation}

From point 5. it follows, that $<1>_{\beta}=Sp[\rho(\beta)]$.
Therefore for any scalar quantity $f$ we have $<f>_{\beta}=f
Sp[\rho(\beta)]$. In particular, the mean value
$<[x_{\mu},p_{\nu}]>_{\beta}$ is equal to
\begin{equation}\label{U15}
<[x_{\mu},p_{\nu}]>_{\beta}= i\hbar\delta_{\mu,\nu}
Sp[\rho(\beta)]
\end{equation}
We will call density matrix the limit
$\lim\limits_{\beta\rightarrow 0}\rho(\beta)=\rho$. It is evident,
that in the limit $\beta\rightarrow 0$ we turn to QM. Here we
would like to verify, that two cases described above correspond to
the meanings of $\beta$. In the first case when $\beta$ is near to
1/4. In the second one when it is near to zero.
\\
From the definitions given above it follows that
$<(j><j)>_{\beta}=\omega_{j}(\beta)$. From which the condition of
completeness on $\beta$ is
\\$<(\sum_{i}|i><i|)>_{\beta}=<1>_{\beta}=Sp[\rho(\beta)]$. The
norm of any vector $|\psi>$, assigned to  $\beta$ can be defined
as
\\$<\psi|\psi>_{\beta}=<\psi|(\sum_{i}|i><i|)_{\beta}|\psi>
=<\psi|(1)_{\beta}|\psi>=<\psi|\psi> Sp[\rho(\beta)]$, where
$<\psi|\psi>$ is the norm in QM, or in other words when
$\beta\rightarrow 0$. By analogy, for probabilistic interpretation
the same situation takes place in the described theory, but only
changing $\rho$ by $\rho(\beta)$.
\\

\renewcommand{\theenumi}{\Roman{enumi}}
\renewcommand{\labelenumi}{\theenumi.}
\renewcommand{\labelenumii}{\theenumii.}

Some remarks:

\begin{enumerate}
\item The considered above limit covers at the same time
Quantum and Classical Mechanics. Indeed, since
$\beta=l_{min}^{2}/x^{2 }=G \hbar/c^3 x^{2 }$, so we obtain:
\begin{enumerate}
\item $(\hbar \neq 0,x\rightarrow
\infty)\Rightarrow(\beta\rightarrow 0)$ for QM;
\item $(\hbar\rightarrow 0,x\rightarrow
\infty)\Rightarrow(\beta\rightarrow 0)$ for Classical Mechanics;
\end{enumerate}
\item The parameter of deformation $\beta$
should take the meaning $0<\beta\leq1$. However, as we can see
from (\ref{U14}), and as it was indicated in the section 2,
$Sp[\rho(\beta)]$ is well defined only for $0<\beta\leq1/4$.That
is if $x=il_{min}$ and $i\geq 2$ then, there is not any problem.
 At the very point with fundamental
length $x=l_{min}\sim L_{p}$ there is a singularity, which is
connected with the appearance of the complex value  of
$Sp[\rho(\beta)]$, or in other words it is connected with the
impossibility of obtain a diagonalized density pro-matrix at this
point over the field of real numbers. For this reason definition 1
at the initial point do not has any sense.
\item We have to consider the question about solutions
(\ref{U13}). For instance, one of the solutions (\ref{U13}), at
least at first order on $\beta$ is
$\rho^{*}(\beta)=\sum_{i}\alpha_{i} exp(-\beta)|i><i|$, where all
$\alpha_{i}>0$ do not depend on $\beta$  and their sum is equal to
1, that is $Sp[\rho^{*}(\beta)]=exp(-\beta)$. Indeed, we can easy
verify that
\begin{equation}\label{U15}
Sp[\rho^{*}(\beta)]-Sp^{2}[\rho^{*}(\beta)]=\beta+O(\beta^{2}).
\end{equation}
 Note that in the momentum
representation $\beta=p^{2}/p^{2}_{max}$, where $p_{max}\sim
p_{pl}$ and $p_{pl}$ is the Planck's momentum. When present in
matrix elements, $exp(-\beta)$ can damp the contribution of great
momenta in a perturbation theory.
\item It is clear, that in the proposed description of
states, which have a probability equal to 1, or in others words
pure states can appear only in the limit $\beta\rightarrow 0$, or
when all states $\omega_{i}(\beta)$ except one of them are equal
to zero, or when they tend to zero at this limit.
\item We suppose, that all definitions concerning
density matrix can be transferred to the described above
deformation of Quantum Mechanics (QMFL) changing the density
matrix $\rho$ by the density pro-matrix $\rho(\beta)$ and turning
then to the low energy limit $\beta\rightarrow 0$. In particular,
for statistical entropy we have
\begin{equation}\label{U16}
S_{\beta}=-Sp[\rho(\beta)\ln(\rho(\beta))].
\end{equation}
The quantity $S_{\beta}$, evidently never is equal to zero, since
$\ln(\rho(\beta))\neq 0$ and, therefore $S_{\beta}$ may be equal
to zero only at the limit $\beta\rightarrow 0$.
\end{enumerate}

\renewcommand{\theenumi}{\arabic{enumi}}

\section{Some Implications}

\begin{enumerate}
\item If we carry out a measurement in a defined scale,we cannot
consider a density pro-matrix, density pro-matrix with a
precision, which exceed some limit of order $\sim10^{-66+2n}$ ,
where $10^{-n}$ is the scale in which the measurement is carried
out. In most of the known cases this precision is quite enough for
considering density pro-matrix the density matrix. However, at the
Planck scale, where Quantum Gravity effects cannot be neglected
and energy is of the Planck order the difference between $\rho(x)$
and $\widehat{\rho}$  have to be considered.

\item At the Planck scale the notion of wave function of the
Universe, introduced by J.A. Wheeler and B. deWitt \cite{r9} does
not work and in this case quantum gravitation effects can be
described only with the help of density pro-matrix $\rho$.

\item Since density pro-matrix $\rho$ depends on the scale in which
the measurement is carried out, so the evolution of the Universe
within inflation model paradigm \cite{r10} is not an unitary
process, because, otherwise the probability $p_{i}$ would be
conserved.

\item As density pro-matrix $\rho$ for a pure state does not exist,
so $\ln\rho \neq 0$ and statistical entropy
$S=-Sp[\rho\ln\rho]\neq0$ is never equal to zero at the Planck
scale, and condition $S=-Sp[\rho\ln\rho]=0$ can be used only with
a certain degree of precision depending on the scale.
\end{enumerate}

\section{On the problem of black holes information paradox}

The obtained above results give us a new approach to the solution
of Hawking problem on coherence (unitary) and information loss in
black holes \cite{r11}. Indeed, the relation (\ref{U12}) describes
the limit transition from density pro-matrix $\rho$ to density
matrix $\widehat{\rho}$ or in others words, from GUR to UR. Is it
possible the inverse transition from density matrix
$\widehat{\rho}$ to density pro-matrix $\rho$ and correspondingly,
from (\ref{U1}) to (\ref{U2}) ? The answer is affirmative. This
transition is possible when matter are absorbed by a black hole if
we consider,that quantum gravity effects are important when we are
trying to describe physical effects in a black hole, as it was do
in (\ref{U2})\cite{r12}. Thereby we have the next symmetric and
equivalent transitions:

 \vspace*{0.25cm} \noindent I.GUR(Big Bang,Origin
Singularity)$\mapsto$ UR $\mapsto$ GUR(Black
Holes,Sin\-gu\-la\-ri\-ties);

\vspace*{0.25cm} \noindent II.Density Pro-Matrix(mix
states)$\mapsto$ Density Matrix(mix states,pure sta\-tes)$\mapsto$
Density Pro-Matrix(mix states).

 \vspace*{0.25cm} In all papers on
coherence and information loss in black holes (for instance,
\cite{r13},\cite{r14}) so far, the authors have handled with the
right side I and II or another words, with the transitions UR
$\mapsto$ GUR(Black Holes,Singularities), Density Matrix(mix
states,pure states)$\mapsto$ Density Pro-Matrix(mix states), which
are non-unitary according to the obtained above results. However
it is evidently, that it is more rightful if we study I and II
completely, in other words, if we add the left sides GUR(Big
Bang,Origin Singularity)$\mapsto$ UR è Density Pro-Matrix(mix
states)$\mapsto$ Density Matrix(mix states,pure states)
correspondingly. Then, starting from Density Pro-Matrix(mix
states) we come back to  Density Pro-Matrix(mix states) and
unitarity and information can be restored. It is necessary to
remark,that for primordial black holes in I and II their middle
part vanishes, since all processes take place in the early
universe, and we obtain the same result

\vspace*{0.25cm} \noindent Density Pro-Matrix(mix
states))$\mapsto$ Density Pro-Matrix(mix states)
 \vspace*{0.25cm}

\noindent Besides that, it is evident, that the appearance of a
space-time singularity, except of the initial singularity in
Classical Theory, in Quantum Theory  means the transition from UR
to GUR or from (\ref{U1}) to (\ref{U2}). In the case of initial
singularity we have Quantum Theory with GUR from the very
beginning.

\section{Conclusion}

As it was noted in  \cite{r15} all known approaches to justify
Quantum Gravity one way or another lead to the notion of
fundamental length. Besides that GUR (\ref{U2}), which as well
lead to that notion are well described within the inflation model
\cite{r16}. Therefore to understand physics at the Planck scale
without these notions, apparently is not possible. Besides that,
it is necessary to consider one more aspect of this problem. As it
was noted in \cite{r17} , when a new physical theory is created,
it implies the introduction of a new parameter and the deformation
of precedent theory by this parameter. All these deformation
parameters are in their essence fundamental constants: $G$,$c$ and
$\hbar$ (more exactly in \cite{r17} $c$ è $\hbar$ instead of $c$,
$1/c$ is used). It is possible to join some these parameters in an
unique theory. For example $G$ and $c$ in the General Theory of
the relativity, $c$ and $\hbar$ in QFT. In \cite{r17} the next
question was formulated : what could be the theory , which
contains all three fundamental constants, or in other words all
three deformation parameters? From all these follows the question
in \cite{r17} can be revised: is the theory with fundamental
length, the theory which by definition contains all three
fundamental parameters $L_{p}=\surd\frac{G\hbar}{c^3}$ by
definition ?. In \cite{r17} the limit transition from one Quantum
Mechanics to another, described as $L_{p}^2/x^2\rightarrow 0$ can
be understood as $L_{p}\rightarrow 0$, that in the considered case
corresponds either $G\rightarrow0$,$c\rightarrow\infty$ (in the
case of Quantum Mechanics), or $G\rightarrow0$,$c$ tend to a
finite quantity (in the case of Relativistic Quantum theory).




\begin{thebibliography}{99}
%
%
\bibitem{r1}
W.Heisenberg,Uber den anschaulichen Inhalt der
quantentheoretischen Kinematik und Mechanik,
 Zeitsch.fur Phys,43(1927)172
%
%
\bibitem{r2}
G.Veneziano,A stringly nature needs just two constant
Europhys.Lett.2(1986)199;D.Amati,M.Ciafaloni
 and G.Veneziano,Can spacetime be probed below the
 string size? Phys.Lett.B216(1989)41;
E.Witten, Reflections on the Fate of Spacetime
Phys.Today,49(1996)24
%
%
\bibitem{r3}
R.J.Adler and D.I.Santiago,On Gravity and the Uncertainty
Principle, Mod.Phys.Lett.A14(1999)1371[gr-qc/9904026]
%
%
\bibitem{r4}
M.Maggiore, A Generalized Uncertainty Principle in Quantum Gravity
Phys.Lett.B304(1993)65,[hep-th/9301067]
%
%
\bibitem{r5}
D.V.Ahluwalia,Wave-Particle duality at the Planck scale: Freezing
of neutrino oscillations Phys.Lett. A275 (2000)31,
[gr-qc/0002005];Interface of Gravitational and Quantum Realms
Mod.Phys.Lett. A17(2002)1135,[gr-qc/0205121]; M.Maggiore,Quantum
Groups,Gravity and Generalized Uncertainty Principle
Phys.Rev.D49(1994)5182,[hep-th/9305163]; The algebraic structure
of the generalized uncertainty principle
Phys.Lett.B319(1993)83,[hep-th/9309034];S.Capozziello,G.Lambiase
and G.Scarpetta, The Generalized Uncertainty Principle from
Quantum Geometry [gr-qc/9910017]
%
%
\bibitem{r6}
Y.J.Ng, H.van Dam,Measuring the Foaminess of Space-Time with
Gravity-Wave Interferometers,Found.Phys.30(2000)795,
[gr-qc/9906003]
%
%
\bibitem{r7}
Y.J.Ng, H.van Dam, On Wigner's clock and the detectability
space-time foam with gravitational-wave interferometers
Phys.Lett.B477(2000)429,[gr-qc/9911054]
%
%
\bibitem{r8}
J.C.Baez, S.J.Olson,Uncertainty in Measurment of
Distance,[gr-qc/0201030]
%
%
\bibitem{r9}
J.A.Wheeler,in Battele Rencontres,ed. by C.DeWitt and J.A. Wheeler
(Benjamen,NY,1968)123; B.DeWitt,Quantum Thery Gravity I.The
Canonical Theory, Phys.Rev.160(1967)1113.
%
%
\bibitem{r10}
A.H.Guth,Inflation and EternaL Inflation,[astro-ph/0002156]
%
%
\bibitem{r11}
S.Hawking,Breakdown of Predictability in Gravitational Collapse,
Phys.Rev.D14(1976)2460
%
%
\bibitem{r12}
R.Adler,P.Chen and D.Santiago,The Generalised Uncertainty
Principle and Black Hole
Remnants,Gen.Rel.Grav.33(2001)2101,[gr-qc/0106080]; P.Chen and
R.Adler, Black Hole Remnants and Dark Matter,[gr-qc/0205106]
\bibitem{r13}
S.Giddings,The Black Hole Information Paradox,[hep-th/9508151]
%
%
\bibitem{r14}
A.Strominger, Les Houches Lectures on Black Holes,
[hep-th/9501071]
%
%
\bibitem{r15}
L.Garay,Quantum Gravity and Minimum Length
Int.J.Mod.Phys.A.v.A10(1995)145
%
%
\bibitem{r16}
S.F.Hassan and M.S.Martin, Trans-Plancian Effects in Inflationary
Cosmology and Modified Uncertainty Principle, [hep-th/0204110]
%
%
\bibitem{r17}
L.Faddeev, Mathematical View on Evolution of Physics, Priroda
5(1989)11
%
%
\bibitem{r18}
A.Kempf,G.Mangano,R.B.Mann,Hilbert Space Representation of the
Minimal Length Uncertainty
Relation,Phys.Rev.D52(1995)1108[hep-th/9412167]
%
%
\end{thebibliography}
\end{document}